\documentclass{article}    
\usepackage{amsmath}
\usepackage{amssymb}
\usepackage{graphicx}

\newtheorem{defi}{Definition}
\newtheorem{prop}{Proposition}

\newtheorem{pb}{Problem}

\newenvironment{proof}[1][Proof]{\begin{trivlist}
    \item[\hskip \labelsep {\bfseries #1}]}{\end{trivlist}}

\newcommand{\AAA}{A}  
\newcommand{\DD}{D}   
\newcommand{\LL}{L}   

\newcommand{\FF}{\mathcal{F}} 
\newcommand{\FFc}{\mathcal{F}_0} 
\newcommand{\GG}{\mathcal{G}}    

\newcommand{\HH}{\mathcal{H}} 

\newcommand{\XX}{\mathcal{X}}   
\newcommand{\Eset}{\mathcal{E}} 

\newcommand{\ee}{e}    

\newcommand{\hh}{h}
\newcommand{\vv}{v}

\newcommand{\dR}{\mathbb{R}}
\newcommand{\dN}{\mathbb{N}}
\newcommand{\II}{{\bf 1}}
\title{Graph-driven features extraction from microarray data} 
\author{Jean-Philippe Vert and Minoru Kanehisa\\
Bioinformatics Center\\
 Institute for Chemical Research\\
 Kyoto University\\
 Uji, Kyoto 611-0011, Japan\\
\texttt{Jean-Philippe.Vert@mines.org}\\
\texttt{kanehisa@kuicr.kyoto-u.ac.jp}}

\date{June 15, 2002}

\begin{document}
\maketitle
\begin{abstract}
Gene function prediction from microarray data is a first step toward better understanding the machinery of the
 cell from relatively cheap and easy-to-produce data. In this paper we investigate
 whether the knowledge of many metabolic pathways and their catalyzing
 enzymes accumulated over the years can help improve the performance of
 classifiers for this problem.

The complex network of known biochemical reactions in the cell results
 in a representation where genes are nodes of a graph.
Formulating the problem as a graph-driven features extraction problem,
based on the simple idea that relevant features are likely to
exhibit correlation with respect to the topology of the graph, we
end up with an algorithm which involves encoding the network and the
set of expression profiles into kernel functions, and performing a
regularized form of canonical correlation analysis in the corresponding
reproducible kernel Hilbert spaces.

Function prediction experiments for the genes of the yeast {\em
 S. Cerevisiae} validate this approach by showing a consistent increase
 in performance when a state-of-the-art classifier uses the vector of
 features instead of the original expression profile to predict the
 functional class of a gene.

\end{abstract}


\section{Introduction}  

Following the near completion of many genome sequencing projects and the
identification of genes coding for proteins in these genomes, the
research paradigm is shifting toward a better understanding of the
functions of the genes and their interactions. This discipline, broadly
called {\em functional genomics} is expected to provide new insights
into the machinery of the cell and suggest new therapeutic targets by
better focusing on the precise molecules or processes responsible for a
given disease.

Functional genomics has been boosted since the mid 1990's by the
introduction of the DNA microarray technology \cite{sche95,brow00b},
which enables the monitoring of the quantity of messenger
RNA (mRNA) present in a cell for several thousands genes simultaneously,
at a given instant. As
mRNA is the intermediate molecule between the blueprint of a
protein on the DNA strand and the protein itself, it is expected that
the quantity of mRNA reflects the quantity of the protein itself, and
that variations in the quantity of mRNA when a cell is confronted to
various experimental conditions reflects the genetic regulation
process. Consequently functional characterization of a protein from its
expression profile as measured by several microarray hybridation
experiments is supposed to be possible to some extent, and initial
experiments confirmed that many genes with similar function yield
similar expression patterns \cite{eise98}. As data accumulate the
incentive to develop precise methods to assign functions to genes from
expression profiles increases.

Proteins can have many structural or functional roles. In particular
proteins known as enzymes catalyze chemical reactions which enable
cells to acquire energy and materials from its environment, and to
utilize them to maintain their own biochemical network. Decades of
careful experiments have helped characterize many reactions taking place
in the cell together with some of the genes playing a role in their
control, and this information has now been integrated into several
databases including WIT \cite{over00} or KEGG \cite{kane02}. Such
databases provide a view of the set of proteins as the nodes of a large
and complex network, where two genes are linked when they catalyze two
successive reactions.

The question motivating this paper is whether this network can help
improve the performance of function prediction algorithms based on
microarray data only. To this end we propose a graph-driven feature
extraction process from the expression profiles, based on the idea that
patterns of expression which correspond to actual biological events,
such as the activation of a series of chemical reactions forming a {\em
chemical pathway}, are likely to be shared by genes close to each other
with respect to the network topology. Translating this idea
mathematically we end up with a features extraction process equivalent to
performing a generalization of canonical correlation analysis (CCA) between the representations
of the genes in two different reproducing kernel Hilbert spaces, defined
respectively by a diffusion kernel \cite{kond02} on the gene graph and
by a linear kernel on the expression profiles. The CCA can be performed
in these RKHS using the kernel-CCA algorithm presented in \cite{bach01}.

Relationships between expression profiles and biochemical pathways have
been subject to much investigation in the recent years. As
microarray data are much cheaper to produce than precise pathway data,
pathway reconstruction or validation from expression data has been
attracting much attention since the availability of public microarray data
\cite{frie00,akut00}. Extraction of co-clusters, i.e., clusters of
genes in the network which have similar expression has also been
investigated recently \cite{naka01,hani02}. On the technical point of
view the integration of several sources of data has been investigated 
with different approaches, e.g., combining expression data and genomic
location information in a Bayesian framework \cite{hart02}, combining
expression data with phylogenetic profiles by kernel operations
\cite{pavl01}, or defining distances between genes by combining
distances measured from different data types \cite{marc99}.

This paper is organized as follows. Section \ref{sec:pbdef} translates
mathematically the feature extraction problem and contains basic notations
and definitions, followed by a short review of some properties of RKHS
relevant for our purpose in Section \ref{sec:RKHS}. Sections
\ref{sec:net} and \ref{sec:relevance} describe respectively how two
important properties of features can be expressed in
terms of norms in RKHS, and Section \ref{sec:extraction} describes the
feature extraction process. Experimental results are presented in
Section \ref{sec:experiment}, followed by a discussion in Section \ref{sec:discussion}.

\section{Problem definition}\label{sec:pbdef}

\subsection{Setting and notations}

Before focusing on expression profiles and biochemical pathways, we first
formulate in a more abstract way the problem we are dealing with. 
The set of genes is represented by a finite set $\XX$ of cardinality $|\XX| = n$, where each element $x \in \XX$ represents a gene. The information provided
by the microarray experiments and the pathway database are
represented respectively as:
\begin{itemize}
\item a mapping
$
\ee : \XX \rightarrow \dR^p,
$ 
where $\ee(x)$ is the expression profile for the gene $x$, for any $x$ in $\XX$, and $p$ is the number of measurements available. In the sequel we assume that the profiles have been centered, i.e.:
\begin{equation}\label{eqn:centered}
\sum_{x \in \XX} e(x) = 0.
\end{equation}
\item A simple graph $\Gamma = (\XX,\Eset)$ (without loops and multiple
      edges) whose vertices are the genes $\XX$ and
      whose edges $\Eset$ represent the links between genes, as extracted from the biochemical pathway database.
\end{itemize}

The notation $x \sim y$ for any $(x,y) \in \XX^2$ means that there is an edge between $x$ and $y$, i.e., $\{x,y\} \in \Eset$. 
Our goal in the sequel is to use the graph
$\Gamma$ in order to extract features from the expression profiles $\ee$ relevant for the
functional classification of the genes. In this context we formally define a feature
to be a mapping $f:\XX \rightarrow \dR$, and we denote by $\FF = \dR^{\XX}$
the set of possible features. The set of centered features is denoted by
$\FFc = \left\{f \in \FF : \sum_{x \in \XX}f(x) = 0\right\}$. For any
feature $f \in \FF$ the same notation is used to represent the
$n$-dimensional vector $f = \left(f(x)\right)_{x \in \XX}$ indexed by
the elements of $\XX$, and $f'$ denotes its transpose. The constant unit
vector is denoted $\II = (1,\ldots,1)$.

\subsection{Feature relevance}\label{subsec:relevance}
Features can be derived from the mapping $\ee$. As an example,
projecting $\ee$ to a given direction $v \in \dR^p$ gives the feature $f_{\ee,v}$
defined for any $x$ in $\XX$ by:
\begin{equation}\label{eqn:linfeat}
f_{\ee,v}(x) = \vv'\ee(x).
\end{equation}
 If $v$
represents a particular expression pattern, then $f_{\ee,v}$ quantifies
how each gene correlates with this pattern. In this paper we
restrict ourselves to such linear features, and denote by $\GG =
\{f_{e,v} , v \in \dR^p\} \subset \FF$ the set of linear
features. Observe that by hypothesis (\ref{eqn:centered}), each linear
feature is also centered by (\ref{eqn:linfeat}), i.e., $\GG \subset \FFc$.

Biological events such as synthesis of new molecules or transport of
substrates usually require the coordinated actions
of many proteins. Genes encoding such proteins are therefore likely to
share particular patterns of expression over different experimental
conditions, e.g. simultaneous overexpression or inhibition. A vector $v
\in \dR^p$ representing this pattern should therefore be particularly
correlated (positively or negatively) with the genes participating in
the biological process. As a result, linear features $f_{e,v}$
corresponding to biologically relevant patterns $v
\in \dR^d$ are more likely to have a larger variance than those
corresponding to patterns unrelated to any biological event, where the
variance is defined by:
\begin{equation}\label{eqn:variation}
\forall f_{e,v} \in \GG, \quad V(f_{e,v}) = \frac{\sum_{x \in \XX} f_{e,v}(x)^2}{||v||^2} .
\end{equation}

On the other extreme a pattern $v \in \dR^p$ orthogonal to all profiles
leads to a feature $f_{e,v}$ with null variance, and is clearly unlikely
to be related to any biological process requiring gene expression. It
follows that the variance (\ref{eqn:variation}) captured by a feature is
a first indicator of its biological pertinence. In order to prevent
confusion with other criteria in the sequel, we will call a 
feature {\em relevant} if it captures much variations between expression
profiles in the sense of
(\ref{eqn:variation}), and {\em irrelevant} otherwise. The reader can
observe that searching for the most relevant features can be done by
performing a principal component analysis (PCA) \cite{joll86} of the
profiles, the first principal components corresponding to the most
relevant features; however we now show that relevance is not the only
criterion which can be used to select features.

\subsection{Feature smoothness}\label{sec:smoothness}

Relevance as defined in Section \ref{subsec:relevance} is an intrinsic
properties of the set of profiles, as it is defined in terms of
variation captured, and no other information about the relationships
between genes is used.

Independently of any microarray experiment, many
 metabolic pathways
have been experimentally characterized over the years. These collections of chemical reactions involve proteins as enzymes, whose presence or absence plays a major role in monitoring the reaction.
Actual biological event usually involve series of such reactions, also called {\em pathways}. Genes involved in consecutive reactions of pathways
are likely to share particular patterns of expression, corresponding to the
activation or not of the corresponding pathway.

As a result a pattern $v \in \dR^p$ which corresponds to a true
biological event, such as the activation or inhibition of a
pathway, is likely to be shared by clusters of genes in the
graph of genes where two genes are linked if they participate in
consecutive reactions. On a more global scale, such a feature is more
likely to vary smoothly on the graph of genes, in the sense that
variations between linked genes be as small as possible, than a noisy
pattern unrelated to any biochemical event which would not exhibit any
particular correlation between genes linked to each other in the graph.

Such features are called {\em smooth} in the sequel, by opposition
to {\em rugged} features which vary a lot with respect to the graph topology.
These notions are formalized and
quantified in terms of a norm in a Hilbert space in Section
\ref{sec:net}, but before developing these technicalities we can already
sketch
a feature extraction process based on this intuitive definition.

\subsection{Problem formulation}\label{sec:pbform}

From the discussions in Sections \ref{subsec:relevance} and
\ref{sec:smoothness} two criteria appear to characterize ``good'' candidate
features : their relevance on the one
hand (Section \ref{subsec:relevance}) based on a statistical
analysis of the set of profiles, and their smoothness on the other hand
(Section \ref{sec:smoothness}) which results from the analysis of the
variations of the feature with respect to the topology of the graph of
genes. 

Good candidate features are smooth and relevant in the same
time. These two properties are however not always correlated: it might
be possible to find many relevant but rugged features, as well as
smooth but irrelevant features. A reasonable approach to extract
meaningful features is therefore to try to find a compromise between
these two criteria, and to extract features which are as smooth and
relevant in the same time as possible.

Although this statement can be translated mathematically in many
different ways, we investigate in the sequel the following formulation:

\begin{pb}\label{pb:init}
Extract pairs of features $(f_1,f_2) \in \FF_0 \times \GG$ such that:
\begin{itemize}
\item $f_1$ be smooth,
\item $f_2$ be relevant,
\item $f_1$ and $f_2$ be correlated.
\end{itemize}
\end{pb}

These three goals are usually contradictory and a trade-off must be
found between them.
Observe that if either the smoothness or the relevance conditions are
removed, the problem is likely to be ill-posed. For instance, if the smoothness requirement
is removed then any relevant feature $f_2$ is perfectly correlated with
itself; on the other hand if the relevance conditions disappears then
many smooth features $f_1$ can probably be correlated with linear
features which are not necessarily relevant (this possibility increases
when the dimension $p$ of the profiles increases, as the set of linear
features increases too).

Let us now formulate Problem \ref{pb:init} mathematically.
The correlation between any two centered features $(f_1,f_2) \in
\FFc^2$ is equal to:
\begin{equation}\label{eqn:corr}
c(f_1,f_2) = \frac{f_1'f_2}{\sqrt{f_1'f_1}\sqrt{f_2'f_2}} .
\end{equation}
As already mentioned the maximization of $c(f_1,f_2)$ over $\FF_0 \times
\GG$ is an ill-posed problem.

Suppose we can define a smoothness functional $\hh_1: \FF \rightarrow
\dR^+$ for any feature, and a relevance functional $\hh_2: \GG
\rightarrow \dR^+$ for linear features, in such a way that lower values
of the functional $\hh_1$ (resp. $\hh_2$) corresponds to smoother
(resp. more relevant) features. Then one way to formalize the trade-off
between correlation and relevance / smoothness is to solve the following maximization problem:
\begin{equation}\label{eqn:pb}
\max_{(f_1,f_2) \in \FF_0 \times \GG} \frac{f_1'f_2}{\sqrt{f_1'f_1 + \delta \hh_1(f_1)}\sqrt{f_2'f_2 + \delta \hh_2(f_2)}} ,
\end{equation}
where $\delta$ is a regularization parameter. When $\delta = 0$ we
recover the ill-posed problem of maximizing the correlation
(\ref{eqn:corr}), and the larger $\delta$ the smoother (resp. the more
relevant) the feature $f_1$ (resp. $f_2$) which solves (\ref{eqn:pb}). As
a result, a solution $(f_1,f_2)$ of (\ref{eqn:pb}) is a reasonable
solution to Problem \ref{pb:init}, with $\delta$ controlling the
trade-off between correlation on the one hand, smoothness and relevance
on the other hand. 

Equation (\ref{eqn:pb}) is therefore the problem we consider
is the sequel. In order to
solve it we need to 1) express the relevance and smoothness functional
$h_1$ and $h_2$ mathematically and 2) solve the maximization problem
(\ref{eqn:pb}) with these functionals.
These two steps are not independent. In particular there is an incentive
to express mathematically $h_1$ and $h_2$ in such a way that
(\ref{eqn:pb}) be computationally solvable. 

If $f_1$ and $f_2$ were restricted to be linear functionals obtained by
projecting two different vector representations of the genes on
particular directions, then the maximization of (\ref{eqn:corr}) would
be the exactly the first {\em canonical correlation} between $f_1$ and
$f_2$ \cite{hote36}, as obtained by classical canonical correlation
analysis (CCA). Linear algebra algorithms involving eigenvector
decomposition exist to perform CCA. However $f_1$ is not restricted to be a linear
feature, and (\ref{eqn:corr}) is consequently ill-posed.

Formulated as (\ref{eqn:pb}), however, we recover a slight
generalization of CCA introduced in
\cite{bach01} and called {\em kernel-CCA}. More precisely, kernel-CCA is
formulated as:
\begin{equation}\label{eqn:kernelCCA}
\max_{(f_1,f_2) \in \HH_1 \times \HH_2} \frac{f_1'f_2}{\sqrt{f_1'f_1 + \delta ||f_1||_{\HH_1}}\sqrt{f_2'f_2 + \delta ||f_2||_{\HH_2}}} ,
\end{equation}
where $\HH_1$ and $\HH_2$ are two reproducible kernel Hilbert spaces (see Section
\ref{sec:RKHS}) on the space $\XX$. Problem (\ref{eqn:kernelCCA}) is
equivalent to a generalized eigenvalue problem \cite{bach01} and can be
solved iteratively to extract several pairs of features (see Section
\ref{sec:featext}).

In order to use the algorithm of \cite{bach01} we therefore need to restate
(\ref{eqn:pb}) in terms of optimization in RKHS like (\ref{eqn:kernelCCA}).
This involves 1) expressing $\FF_0$ as a RKHS whose norm is a smoothness
functional (Section \ref{sec:smoothness}), 2) expressing $\GG$ as a RKHS
whose norm is a relevance functional (Section \ref{sec:relevance}), and
3) solving the resulting problem (\ref{eqn:kernelCCA}).

\section{Reproducing kernel Hilbert space}\label{sec:RKHS}
Before carrying out the program sketched in Section \ref{sec:pbform} we
first recall some definitions and basic properties of RKHS in order to
make this paper as self-contained as possible. Good introductions on
RKHS can be found in \cite{aron50,sait88,wahb90,scho02} from which we
borrow most of the materials presented in this section.

\subsection{Basic definitions}
Let $\XX$ be a set (which we don't necessarily assume to be finite in
this section), and $K: \XX \rightarrow \dR$ a symmetric positive definite
function, in the sense that for every $l \in \dN$ and
$(x_1,\ldots,x_l) \in \XX^l$ the $l \times l$ Gram matrix $K_{i,j} =
K(x_i,x_j)$ be positive semidefinite.

Then it is known that the linear span
of set of
functions $\{K(.,x), x \in \XX\} \subset \dR^{\XX}$ can be
completed into a Hilbert space $\HH \subset \dR^{\XX}$ which satisfies
the following ``reproducing property'':
\begin{equation}\label{eqn:repro}
\forall (f,x) \in \HH \times \XX, \quad f(x) = \left< K(.,x) , f
\right>_{\HH} ,
\end{equation}
where $<.,.>_{\HH}$ represents the inner product of $\HH$. In
particular, by plugging $f = K(.,x')$ in (\ref{eqn:repro}) we obtain:
\begin{equation}\label{eqn:KK}
\forall  (x,x') \in \XX^2, \quad \left< K(.,x) , K(.,x')\right>_{\HH} =
K(x,x') .
\end{equation}
The Hilbert space $\HH$ is called a {\em reproducing kernel
Hilbert space} \cite{aron50} to emphasize the property (\ref{eqn:repro}).
In order to make this rather abstract result clearer, let us show how
the space $\HH$ can be built when $\XX$ is finite, which is the case of
interest in this paper. 

Let us therefore take $\XX$ to be the finite set of genes, and suppose first
that the $n \times n$ Gram matrix $K_{x,y} = K(x,y)$ for any $(x,y) \in \XX^2$
is {\em positive definite}, i.e., that its eigenvalues are all positive. It
can then be diagonalized as follows:
\begin{equation}\label{eqn:k1}
K = \sum_{i=1}^n \lambda_i \phi_i\phi_i' ,
\end{equation}
where the eigenvalues satisfy $0 < \lambda_1 \leq \ldots \leq
\lambda_n$ and the set $(\phi_1,\ldots,\phi_n) \in \FF^n$ is an associated orthonormal
basis of eigenvectors. 

We can now take the Hilbert space to be $\HH=\FF$, and define the inner product in $\HH$ in terms of the decomposition of any $f \in \HH$ in the basis of eigenvectors:
\begin{equation}\label{eqn:k2}
f = \sum_{i=1}^n a_i \phi_i ,
\end{equation}
as follows:
\begin{equation}\label{eqn:finiteK}
\left< \sum_{i=1}^n a_i \phi_i ,  \sum_{i=1}^n b_i \phi_i \right>_{\HH}
= \sum_{i=1}^n \frac{a_ib_i}{\lambda_i} .
\end{equation}
It is easy to check that the Hilbert space defined by
(\ref{eqn:finiteK}) satisfies the reproducing property
(\ref{eqn:repro}), and is therefore a RKHS associated with the kernel
$K(.,.)$.

The columns of the Gram matrix being independent,
any feature $f \in \HH$ can be uniquely represented as follows:
\begin{equation}\label{eqn:dualform}
f(.) = \sum_{x \in \XX} \alpha(x) K(x,.) ,
\end{equation}
or in an equivalent matrix form:
\begin{equation}\label{eqn:primaldual}
f = K\alpha.
\end{equation}
This representation is called the {\em dual} representation of $f$, and
the vector $\alpha = \left(\alpha(x)\right)_{x \in \XX} \in \FF$ is
called the {\em dual coordinate} of $f$.

The dual representation is useful to express the inner
product in the Hilbert space $\HH$.
Indeed, using (\ref{eqn:dualform}) and (\ref{eqn:KK}) it is easy to check that
the inner product between two features $(f,g) \in \FF^2$  with dual
coordinates $(\alpha , \beta) \in \FF^2$ respectively
is given by:
$$
\left<f , g\right>_{\HH} = \sum_{(x,y) \in \XX^2} \alpha(x)
\beta(y) K(x,y) = \alpha' K \beta .
$$

In particular the $\HH$-norm of a feature $f \in \FF$ with dual
coordinates $\alpha \in \FF$ is given by:
\begin{equation}\label{eqn:dualnorm}
|| f ||_{\HH}^2 =   \alpha' K \alpha .
\end{equation}

The inner product in the original space $L^2(\XX)$ can also simply be
expressed with the dual representation: for any $(f,g) \in \FF^2$ with
dual coordinates $(\alpha,\beta)$ respectively we have by
(\ref{eqn:primaldual}) and using the fact that $K$ is symmetric:
$$
f'g = \sum_{x \in \XX} f(x)g(x) = \alpha' K^2 \beta .
$$

In case the kernel $K$ is just positive semidefinite, with $r$ being the
multiplicity of $0$ as eigenvalue, then we can follow the same
construction with the index $i$ ranging from $r+1$ to $n$ in
(\ref{eqn:k1}), (\ref{eqn:k2}) and (\ref{eqn:finiteK}). In that case the
RKHS $\HH$ is the linear span of $\{\phi_{r+1},\ldots,\phi_{n}\}$, of
dimension $n-r$. The dual representation still makes sense but is
defined up to an element of $\{\alpha \in \dR^{\XX}, K\alpha=0\}$.

\subsection{RKHS and smoothness functional}\label{sec:rkhssmooth}

One classical application of the theory of RKHS is regularization theory
to solve ill-posed problems
\cite{tikh77,ivan76,wahb90,giro95}. Indeed it is well known that for
many choices of kernels $K(.,.)$ on continuous spaces $\XX \subset
\dR^N$ the norm in the corresponding RKHS
$||f||_{\HH}$ is intimately related to the smoothness properties of the
functions $f \in \HH$.

The following classical example is relevant for us. Consider a set $\XX
\subset \dR^N$ and a translation-invariant 
kernel of the form $K(x,y) = k(x-y)$ for any $(x,y) \in \XX^2$. Then the RKHS $\HH$ is composed of
the functions $f \in L^2(\XX)$ such that:
\begin{equation}\label{eqn:smooth}
||f||_{\HH} = \int_{\dR^N} \frac{|\hat{f}(\omega)|^2}{\nu(\omega)}
d\omega < \infty,
\end{equation}
where $\hat{f}(\omega)$ is the Fourier transform of $f$ and
$\nu(\omega)$ is the Fourier transform of $k(.)$ \cite{giro95,smol98}. Functionals of the
form (\ref{eqn:smooth}) are known to be smoothness functionals (in which
case smoothness is defined in terms of Fourier transform, i.e., smooth
functions are functions with few energy at high frequency), where the
rate of decrease to zero of $\nu$ controls the smoothness properties of
the function in the RKHS. For example, for the Gaussian radial
basis function $k(x-y) = \exp(-||x-y||^2 / 2\sigma^2)$ the norm in the
RKHS takes the form:
\begin{equation}\label{eqn:gaus}
||f||_{\HH} =  \left(2\pi\sigma^2\right)^{-\frac{p}{2}} \int_{\dR^N}
e^{\frac{\sigma^2}{2}||\omega||^2} |\hat{f}(\omega)|^2 d\omega .
\end{equation}
Equation (\ref{eqn:gaus}) shows that the energy of $f$ at a frequency
$\omega$ should decrease at least as $\exp(- \sigma^2||\omega||^2 / 2)$ for its
$\HH$-norm to be finite. Functions with much energy at high-frequency
have a large norm in $\HH$, which therefore acts as a
smoothness functional. 

We refer the reader to
\cite{tikh77,ivan76,wahb90,giro95} for more details on the connections
between RKHS and smoothness functionals, as well as for applications to
solve ill-posed problems. In the sequel we will adapt these approaches
to discrete spaces $\XX$ in order to fulfill the program sketched in
Section \ref{sec:pbform}

\section{Smoothness functional on a graph}\label{sec:net}

As pointed out in Sections \ref{sec:pbform} our interest is now to
derive a ``smoothness functional'' for features $f \in \FF$ with respect
to the graph $\Gamma$ expressed as a
norm in a RKHS.

\subsection{Fourier transform on graphs}\label{sec:fourier}

Equation (\ref{eqn:smooth}) shows that the norm in a RKHS on a
continuous space associated with a translation-invariant kernel is
defined in terms of Fourier
transform. A natural approach to adapt the construction of smoothing
functional to functions defined on a graph is therefore to adapt the Fourier
transform to that context. As a matter of fact Fourier transforms on graphs have been extensively studied in
spectral graph theory \cite{chun97,moha91,moha97,stad96}, as we now
recall. 

Let $\DD$ be the $n\times n$ diagonal matrix of vertex degrees of the
graph $\Gamma$, i.e., 
$$
\forall (x,y) \in \XX^2, \quad D_{x,y} =
\begin{cases}
 0 &\text{if } x \neq y,\\
 deg(x) &\text{if }x=y,
\end{cases}
$$
where $deg(x)$ is the number of edges involving $x$ in $\Gamma$, and let $\AAA$ be the
adjacency matrix defined by:
$$
\forall (x,y) \in \XX^2, \quad \AAA_{x,y} = 
\begin{cases}
 1 &\text{if there is an edge between $x$ and $y$ in $\Gamma$,}\\
 0 &\text{otherwise .}
\end{cases}
$$
Then the $n \times n$ matrix:
$$
\LL = D - A 
$$
is called the (discrete) {\em Laplacian} of $\Gamma$. The discrete Laplacian $\LL$
is a central concept in spectral graph analysis \cite{moha97}. It shares
many important properties with the familiar differential operator
$$
-\Delta(.) = div(grad(.))
$$
on Riemannian manifolds. It is symmetric, semidefinite positive, and singular. The
eigenvector $(1,\ldots,1)$ belongs to the eigenvalue $\lambda_1 = 0$,
whose multiplicity is equal to the number of connected components of
$\Gamma$.

Let us denote by
$$
0=\lambda_1 \leq \ldots \leq \lambda_n
$$
the eigenvalues of $\LL$ and $\{\phi_i , i=1,\ldots,n\}$ an orthonormal
set of associated eigenvectors. Just like the Fourier basis functions
are eigenfunctions of the continuous Laplacian on $\dR^N$, the
eigenvectors of $\LL$ can be regarded as a discrete Fourier basis on the
graph $\Gamma$ \cite{stad96}, with frequency increasing with their
eigenvalues. 

Although the term ``frequency'' is not well defined for functionals on a
graph, the reader can get an intuition of the fact that the functions $(\phi_i ,
i=1,\ldots,n)$ ``oscillates'' more and more on the graph as $i$
increases through the following two well-known results:
\begin{itemize}
\item Applying the classical equality \cite{moha97}:
$$
\forall f \in \FF, \quad f' \LL f = \sum_{x \sim y}
\left(f(x) - f(y)\right)^2 ,
$$
to an eigenfunction $\phi$ of $\LL$ with
eigenvalue $\lambda$ gives the following equality:
\begin{equation}\label{eqn:eigfun}
 \sum_{x \sim y} \left(\phi(x) - \phi(y)\right)^2 =
 \lambda .
\end{equation}
Equation (\ref{eqn:eigfun}) confirms that the larger $\lambda$, the more
the associated eigenfunction varies between adjacent vertices of the
graph.

\item An other classical result concerns the number of maximal connected
      components of the graph where a feature has a constant sign. The
      first eigenfunction being constant, it has only one such
      component, namely the whole graph. For the other eigenfunctions,
      the discrete nodal domain theorem
which translate Courant's famous nodal theorem for elliptic operators on
Riemannian manifolds \cite{chav84}  to the discrete
settings \cite{verd93,frie93,hols96,davi01} states that the
number of maximal connected subsets of $\XX$ where $\phi_i$ does not
change sign is equal to $i$ in the case where all eigenvalues have multiplicity $1$ (see a more general statement in
\cite{davi01}). Together with the fact that each eigenfunction $\phi_i$
for $i>1$ has zero mean (because it is orthogonal to the constant
function $\phi_1$) this shows that $\phi_i$ ``oscillates'' more and more
      on the graph, in the sense that it changes sign more and more
      often as
      $i$ increases.
\end{itemize}

By similarity with the continuous case the basis
$(\phi_i)_{i=1,\ldots,n}$ is called a {\em Fourier basis}, higher eigenvalues
corresponding to higher frequencies. Any feature $f \in \FF$ can be
expanded in terms
of this basis:
\begin{equation}\label{eqn:fourier}
f = \sum_{i=1}^n \hat{f}_i \phi_i ,
\end{equation}
where $\hat{f}_i = \phi_i' f$ and $\hat{f} = \left( \hat{f}_1,\ldots,\hat{f}_n \right)$ is called the {\em
discrete Fourier transform} of $f$. This provides a way to analyze
features in the frequency domain, and in particular to measure their
smoothness as we now show.

\subsection{Graph smoothness functional}
The Laplacian matrix $\LL$ is semidefinite positive and can therefore
be used as a Kernel Gram matrix. The multiplicity of $0$ as eigenvalue
is the number of connected components of the graph, and the associated
eigenvectors are the functions constant on each connected
components. Following Section \ref{sec:RKHS} the associated RKHS $\HH$ has
dimension $n-r$ and is made of the set of features with zero mean on
each connected component. By
(\ref{eqn:finiteK}) the
norm of any function $f \in \HH$ is given by:
\begin{equation}\label{eqn:normL}
||f||_{\HH}^2 = \sum_{i=r+1}^m \frac{\hat{f}_i^2}{\lambda_i} ,
\end{equation}
where $\hat{f}$ is the Fourier transform of $f$ (\ref{eqn:fourier}) and
$\lambda$ is the 
ordered set of eigenvalues of $L$.

However, as shown in Section \ref{sec:fourier}, the smoothness of
$\phi_i$ decreases with $i$; because $\lambda_i$ increases with $i$, the
norm (\ref{eqn:normL}) in the 
RKHS associated with the kernel $\LL$ increases with smoothness, and is
therefore a ``ruggedness functional'' instead of a smoothness
functional in the sense defined in Section \ref{sec:RKHS}. To illustrate
this we can observe that:
$$
\forall i \in \{r+1,\ldots,n\}, \quad ||\phi_i||_{\HH} =
\frac{1}{\sqrt{\lambda_i}} ,
$$
hence $||\phi_i||_{\HH}$ decreases with $i$.

Transforming this ruggedness functional into a smoothness functional can be performed by a simple operation on the kernel as follows:

\begin{defi}\label{def:zeta}
For any decreasing mapping $\zeta: \dR^+ \rightarrow \dR^+ \backslash \{0\}$, we define
 the $\zeta$-kernel $K_{\zeta} : \XX^2 \rightarrow \dR$ by:
$$
\forall (x,y) \in \XX^2, \qquad K_{\zeta}(x,y) = \sum_{i=1}^n \zeta(\lambda_i) \phi_i(x) \phi_i(y) ,
$$
where $0 = \lambda_1 \leq \ldots \leq \lambda_n$ are the eigenvalues of
 the graph Laplacian and $(\phi_1,\ldots,\phi_n)$ an associated
 orthonormal Fourier basis.
\end{defi}

The mapping $\zeta$ being assumed to take only positive values, the
matrix $K_{\zeta}$ is definite positive and is therefore a valid
kernel, with associated RKHS $\HH = \FF$. From the  
discussion above it is now clear that: 
\begin{prop}\label{prop:zetaK}
The norm $||.||_{\zeta}$ in the RKHS associated with the kernel
 $K_{\zeta}$ is a smoothing functional, given for any feature $f \in \FF$ with Fourier transform $\hat{f} \in \dR^n$ by:
\begin{equation}\label{eqn:smofunc}
||f||_{\zeta}^2 = \sum_{i=1}^n \frac{\hat{f}_i^2}{\zeta(\lambda_i)} .
\end{equation}
\end{prop}
\begin{proof}
Equation (\ref{eqn:smofunc}) is a direct consequence of Definition
\ref{def:zeta} and (\ref{eqn:finiteK}). The fact that
$||.||_{\zeta}$ is a smoothing functional is simply a translation of the
fact that $\zeta(\lambda_i)$ decreases with $i$, hence the relative
contribution of the Fourier components in (\ref{eqn:smofunc}) increases
with their frequency.
\end{proof}

Proposition \ref{prop:zetaK} shows that the smoothness functional
associated with a function $\zeta$ is
controlled by its rate of decrease to $0$. An example of valid $\zeta$
function with rapid decay is the following:
\begin{equation}\label{eqn:exp}
\forall x \in \dR^+, \quad \zeta(x) = e^{-\tau x},
\end{equation}
where $\tau$ is a parameter. In that case we recover the {\em diffusion kernel} introduced and
discussed in \cite{kond02}. The authors of this paper show that
the diffusion kernel shares many properties with the continuous Gaussian
kernel $K(x,y) = \exp(-||x-y||^2/2\sigma^2)$ on $\dR^p$, and can therefore be
considered as its discrete version.

Combining (\ref{eqn:smofunc}) and (\ref{eqn:exp}) we obtain that the norm in the RKHS associated with the
diffusion kernel is given by:
\begin{equation}\label{eqn:normdiff}
\forall f \in \FF, \qquad ||f||_{\zeta} =  \sum_{i=1}^n e^{\tau \lambda_i}\hat{f}_i^2 ,
\end{equation}
hence the high frequency energy of $f$ is strongly penalized by this
kernel, and the penalization increases with the parameter $\tau$.

Before continuing we should observe that in concrete applications the computation of the kernel $K_{\zeta}$ for
 a given $\zeta$ can be performed by diagonalizing the Laplacian matrix
 as:
$$
\LL = \Phi' \Lambda \Phi ,
$$
where $\Lambda$ is a diagonal matrix with diagonal element
 $\Lambda_{i,i} = \lambda_i$, and computing:
$$
K_{\zeta} = \Phi' \zeta(\Lambda) \Phi ,
$$
where $\zeta(\Lambda)$ is a diagonal matrix with diagonal element
 $\zeta(\Lambda)_{i,i} = \zeta(\lambda_i)$. We can also observe that the
 diffusion kernel can be written using the matrix exponential as:
$$
K_{\zeta} = e^{-\tau L} .
$$

Although other choices of $\zeta$ lead to other kernels, discussing them
would be beyond the scope of this paper so we will restrict ourselves to
using the diffusion kernel as a smoothing functional in the sequel. The
conclusion of this section is that by using the diffusion kernel we can build a RKHS $\HH = \FF$
whose norm $||.||_{\HH}$ is a smoothness functional.

\section{Relevance functional}\label{sec:relevance}

Let us now consider the problem of defining a relevance
functional. First observe that any direction $v \in \dR^p$ with
orthogonal projection $v_0$ on the linear span of $\{e(x), x \in \XX\}$
satisfies $f_{e,v} = f_{e,v_0}$. As a result the search of linear
features $f_{e,v}$ can be restricted to directions belonging to this
linear span, which can be parametrized as:
\begin{equation}\label{eqn:beta}
v = \sum_{x \in \XX} \beta(x) e(x) ,
\end{equation}
where $\beta \in \FF$ is called the dual coordinate of $v$ (defined up
to an element of $\{\beta \in \FF, K\beta=0\}$). 

The positive semidefinite Gram matrix $K_{x,y} = e(x)'e(y)$, singular due to the
centering of profiles (\ref{eqn:centered}), defines a RKHS $\HH \subset
\FF$ which consists of features of the form:
\begin{eqnarray*}
f(.) & = & \sum_{x \in \XX} \gamma(x) K(x,.)\\
& = & \sum_{x \in \XX} \gamma(x) e(x)'e(.)\\
& = & \left(\sum_{x \in \XX} \gamma(x) e(x)\right)' e(.) ,
\end{eqnarray*}
where $\gamma \in \FF$. Equation (\ref{eqn:beta}) shows that $\HH$ is
exactly the set of linear features $\GG$, and by (\ref{eqn:dualnorm})
the semi-norm of $\HH$ is given by:
\begin{equation}\label{eqn:relnorm}
\forall f_{e,v} \in \GG, \quad ||f_{e,v}||_{\HH} = \beta' K \beta ,
\end{equation}
where $\beta$ is the dual coordinate of $v$ defined by (\ref{eqn:beta}).

On the other hand, combining (\ref{eqn:linfeat}), (\ref{eqn:variation})
and (\ref{eqn:beta}) shows that the variance of a feature $f_{e,v} \in
\GG$ can be expressed in terms of the dual coordinate $\beta$ of $v$ as
follows:
\begin{eqnarray*}
V(f_{e,v}) & = & \frac{\sum_{x \in \XX} f_{e,v}(x)^2}{||v||^2} \\
& = &  \sum_{x \in \XX}  \frac{\left(v'e(x)\right)^2 }{v'v} \\
& = & \frac{\beta' K^2 \beta}{\beta' K \beta} .
\end{eqnarray*}

From this we see that the larger the ratio between $\beta' K^2 \beta$
and $\beta' K \beta$ the more relevant the feature $f_{e,v}$, where
$v$ has dual coordinates $\beta$.  By observing that $f_{v,e} = K\beta$
and therefore $f_{e,v}'f_{e,v} = \beta' K^2 \beta$, and by
(\ref{eqn:relnorm}) we see that a natural relevance functional to plug
into (\ref{eqn:pb}) in order to
counterbalance the effect of $f_1'f_1$ is the following:
\begin{equation}\label{eqn:relfunc}
h_2(f_{e,v}) = \beta'K\beta = ||f_{e,v}||_{\HH}.
\end{equation}
Indeed the larger $h_2(f_{e,v})$ compared to
$f_{e,v}'f_{e,v}$ the smaller $V(f_{e,v})$, and
therefore the less variation is captured by $f_{e,v}$. The functional
(\ref{eqn:relfunc}) is defined on $\GG$ as the norm of a
RKHS, which was the goal assigned in Section \ref{sec:pbform}.

\section{Extracting smooth correlations}\label{sec:extraction}

\subsection{Dual formulation}

Let us now put together the elements we have developed up to now. In
Section \ref{sec:net} we have shown that any feature $f \in \FF$ can be
represented as:
$$
f = K_1 \alpha ,
$$
where $K_1$ is the diffusion kernel Gram matrix derived from the
Laplacian matrix $L$ by $K_1 = \exp(-\tau L)$, and $\alpha$ is the dual
coordinate vector of $f$ in the corresponding RKHS $\HH_1 = \FF$. Moreover, we defined a
smoothness functional as:
$$
\forall f \in \FF, \quad h_1(f) = ||f||_{\HH_1} = \alpha' K_1 \alpha.
$$

In Section \ref{sec:relevance} we showed that every linear feature
$f_{e,v} \in \GG$ can also be represented in a dual form:
$$
f_{e,v} = K_2 \beta ,
$$
where $K_2$ is the kernel Gram matrix $K_2(x,y) = e(x)'e(y)$ for any
$(x,y) \in \XX^2$ and $\beta$ is the dual coordinate vector in the
corresponding degenerate RKHS $\HH_2 = \GG$. Moreover a relevance functional was defined
as:
$$
\forall v \in \dR^p, \quad h_{e}(f_{e,v}) = ||f||_{\HH_2} = \beta' K_1 \beta.
$$

Plugging these results into (\ref{eqn:pb}) leads to the following formulation of the initial problem in terms of dual coordinates:
\begin{equation}\label{eqn:dualpb}
\max_{(\alpha,\beta) \in \FF^2} \gamma(\alpha,\beta) ,
\end{equation}
with
\begin{equation}\label{eqn:gamma}
 \gamma(\alpha,\beta) =
\frac{\alpha' K_1 K_2 \beta}{\left(\alpha' \left(K_1^2 + \delta
K_1\right) \alpha \right)^{\frac{1}{2}} \left(\beta' \left(K_2^2 + \delta
K_2\right) \beta \right)^{\frac{1}{2}}} .
\end{equation}

Observe that this is the dual formulation of (\ref{eqn:pb}) except that
the optimization is done in $\FF \times \GG$ instead of $\FF_0 \times \GG$.
Moreover, in order keep the
interpretation of $||.||_{\HH_1}$ as a smoothing functional the kernel $K_1$ should not be centered in the feature space, as in usual kernel CCA
\cite{bach01} and kernel PCA \cite{scho99}. As the following Proposition
shows, this is however not a problem because the features whose dual
coordinates maximize (\ref{eqn:dualpb}) are centered anyway, and the
optimization in for $f_1 \in \FF$ is therefore equivalent to the
maximization for $f \in \FF_0$:

\begin{prop}
For any $(\alpha,\beta) \in \FF^2$, let $\alpha_0$ be the dual
 coordinate of the centered version of $f = K_1 \alpha$, i.e.:
$$
\begin{cases}
\exists \epsilon \in \dR, \quad K_1 \alpha_0 = K_1 \alpha + \epsilon\II ,\\
\sum_{x \in \XX} K_1 \alpha_0(x) = 0 .
\end{cases}
$$
Then the following holds:
$$
\gamma(\alpha_0 , \beta) \geq \gamma(\alpha , \beta),
$$
with equality if and only if $\alpha = \alpha_0$. In particular, the
 features whose dual coordinates $\alpha$ and $\beta$ solve
 (\ref{eqn:dualpb}) are centered.
\end{prop}

\begin{proof}
Because the profiles $\{e(x), x \in \XX\}$ are supposed to be centered we have $K_2 \II =0$, and therefore:
$$
\alpha' K_1 K_2 \beta = (\alpha_0' K_1 + \epsilon \II')K_2 \beta = \alpha_0' K_1 K_2\beta .
$$
Let $(\phi_1,\ldots,\phi_n)$ denote an orthonormal Fourier basis, where $\phi_1$ is constant. Then any feature $f=K_1\alpha$ is centered by removing the contribution of $\phi_1$ in its Fourier expansion, i.e., 
$$
f_0 = K_1\alpha_0 = \sum_{i=2}^n \hat{f}_i \phi_i.
$$
As a result we obtain from (\ref{eqn:finiteK}):
\begin{eqnarray*}
\alpha' K_1 \alpha & = & ||f||_{\HH_1}\\
& = & \sum_{i=1}^n \frac{\hat{f}_i^2}{\lambda_i} \\
& \geq & \sum_{i=2}^n \frac{\hat{f}_i^2}{\lambda_i} \\
& = & || K\alpha_0 ||_{\HH_1} \\
& = & \alpha_0 K_1 \alpha_0 ,
\end{eqnarray*}
where the inequality on the third line is an equality if and only if
 $\hat{f}_1=0$, i.e., $f$ is centered. Moreover, using Pythagorean equality
 in $L^2(\XX)$ for the orthogonal vectors $\II$ and $K \alpha_0$ we
 easily get:
\begin{eqnarray*}
\alpha' K_1^2 \alpha & = & ||f||_{L^2(\XX)}\\
& = & ||K_1 \alpha_0 + \epsilon \II ||_{L^2(\XX)}^2 \\
& = &  ||K_1 \alpha_0 ||_{L^2(\XX)}^2 + ||\epsilon \II ||_{L^2(\XX)}^2 \\
& \geq & ||K_1 \alpha_0 ||_{L^2(\XX)}^2\\
& = & \alpha_0' K_1^2 \alpha_0
\end{eqnarray*}
Combining this inequalities with the definition of $\gamma$
 (\ref{eqn:dualpb}) proves the Lemma.
\end{proof}

\subsection{Features extraction}\label{sec:featext}
Stated as (\ref{eqn:dualpb}) the problem is similar to the kernel
canonical correlation problem studied in \cite{bach01}. In particular,
by differentiating with respect to $\alpha$ and $\beta$ we see that
$(\alpha,\beta)$ is a solution of (\ref{eqn:dualpb}) if and only if it
satisfies the following generalized eigenvalue problem:
\begin{equation}\label{eqn:geneig}
\left(
\begin{array}{cc}
0 & K_1K_2\\
K_2K_1 & 0
\end{array}
\right) \left(
\begin{array}{cc}
\alpha\\
\beta
\end{array}
\right) = \rho \left(
\begin{array}{cc}
K_1^2 + \delta K_1& 0\\
0 & K_2^2 + \delta K_2
\end{array}
\right) \left(
\begin{array}{cc}
\alpha\\
\beta
\end{array}
\right)
\end{equation}
with $\rho$ the largest possible. The reader is referred to \cite{bach01}
for details about the derivation of (\ref{eqn:geneig}). Let $\bar{n} =
\min(n,p)$. As pointed out in
this paper solving (\ref{eqn:geneig}) provides a series of pairs of
features:
$$
\left\{\left(\alpha_i,\beta_i\right), i = 1,\ldots,\bar{n}
\right\}
$$
with decreasing values of $\gamma(\alpha_i,\beta_i)$ for which the
gradient $\nabla_{\alpha,\beta} \gamma$ is null, equivalent to the
extraction of successive canonical 
directions with decreasing correlation in classical CCA. The resulting
features $f_{1,i} = K_1 \alpha_i$ and $f_{2,i} = K_2 \beta_i$ are
therefore a set of features likely to have decreasing biological
relevance when $i$ increases, and are the features we propose to
extract in this paper.

The classical way to solve a generalized eigenvalue problem $B\rho =
\lambda C \rho$ is to perform a Cholesky decomposition of $C$ as $C =
E'E$, to define $\mu = E\rho$ and to solve the standard eigenvector
problem $(E')^{-1}BE^{-1} \mu = \lambda \mu$. However the matrix $K_2^2 +
\delta K_2$ is singular so it must be regularized for this approach to
be numerically stable. Following \cite{bach01} this can be done by
adding $\delta^2/4$ on the diagonal, and
observing that: 
$$
K^2 + \delta K + \frac{\delta^2}{4}I = \left(K + \frac{\delta}{2}I\right)^2
,
$$
leads to the following regularized problem:
\begin{equation}\label{eqn:regu}
\left(
\begin{array}{cc}
0 & K_1K_2\\
K_2K_1 & 0
\end{array}
\right) \left(
\begin{array}{cc}
\alpha\\
\beta
\end{array}
\right) = \rho \left(
\begin{array}{cc}
\left(K_1 + \delta' I\right)^2 & 0\\
0 & \left(K_2 + \delta' I\right)^2
\end{array}
\right) \left(
\begin{array}{cc}
\alpha\\
\beta
\end{array}
\right) ,
\end{equation}
where $\delta' = \delta/2$. If $(\alpha,\beta)$ is an generalized
eigenvector solution of (\ref{eqn:regu}) belonging to the generalized
eigenvalue $\rho$, then $(-\alpha,\beta)$ belong to $-\rho$. As a result
the spectrum of (\ref{eqn:regu}) is symmetric :
$(\rho_1,-\rho_1,\ldots,\rho_n,-\rho_n)$ with $\rho_1 \geq \ldots \geq
\rho_n$, $\rho_i = 0$ for $i>p$.

\subsection{Feature extraction process}\label{sec:extpro}

Solving (\ref{eqn:regu}) results in two sets of features $\{K_1\alpha_i,
i=1,\ldots,\bar{n}\}$ and  $\{K_2\beta_i,
i=1,\ldots,\bar{n}\}$. Features of the form $K\alpha_1$ are computed
from the position of the genes in the gene graph, while features of the
form $K_2\beta$ are computed from the expression profiles.

In concrete applications, the position of a still uncharacterized gene
in the gene graph 
is not known, while its expression profile can be measured. As a result
the only way to extract features for such a gene is to use the features
$\left\{K_2\beta_i, i=1,\ldots,\bar{n}\right\}$. These features are
obtained by projecting the expression profiles to the respective directions:
\begin{equation}\label{eqn:eigpro}
v_i = \sum_{x \in \XX} \beta_i(x)e(x), \quad i=1,\ldots,\bar{n}.
\end{equation}
Therefore features can be extracted from any expression profile $e$ by
projections on these directions. We can now summarize a typical
use of the the feature extraction process presented in this paper as
follows:
\begin{itemize}
\item The set of genes $\XX$ is supposed to be the disjoint union of two
      subsets $\XX_1$ and $\XX_2$. Expression profiles are measured for
      all genes, but only genes in $\XX_1$ are present in the gene
      network $\GG = \left(\XX_1,\Eset\right)$. Hence $\XX_1$ is the set of genes which have been
      assigned a precise role in a pathway, while $\XX_2$ is the set of
      uncharacterized genes.
\item Use the set $\XX_1$ to extract features from the set of expression
      profiles $\{e(x), x \in \XX_1\}$ using the graph $\GG$, by solving
      (\ref{eqn:regu}).
\item Derive a set of expression patterns by (\ref{eqn:eigpro}).
\item Extract features from the expression profiles $\{e(x), x \in \XX_2\}$ by projecting them on the derived expression patterns.
\end{itemize}
This process provides a way to replace the expression patterns of an
uncharacterized gene by a vector of features which hopefully are more
biologically relevant than the raw profiles themselves. Any data mining
algorithms, e.g. clustering of
functional classification methods,
can then be applied on this new representation.

\section{Experiments}\label{sec:experiment}

In order to evaluate the relevance of the pathway-driven features
extraction process presented in this paper we performed functional
classification 
experiments with the genes of the yeast {\em Saccharomyces
Cerevisiae}. The main goal of these experiments is to test whether a
state-of-the-art classifier, namely a support vector machine, performs
best by working directly with the expression profiles of the genes, or
by using the vectors of features.

\subsection{Pathway data}
The LIGAND database of chemical compounds and reactions in
biological pathways \cite{goto02,goto98} is part of the Kyoto
Encyclopedia of Genes and Genomes (KEGG) \cite{kane02,kane97}. As of
February 2002 it
consists of a curated set of 3579 metabolic reactions known to take
place in some organisms, together with the substrates involved and the
classification of the catalyzing enzyme as an EC number.To each
reaction are associated one or several EC numbers, and to each EC number
are associated one or several genes of the yeast genome. Using this
information we created a graph of genes by linking two genes whenever
they were assigned two EC number known to catalyze two reactions which
share a common main compound (secondary compounds such as water or ATP
are discarded).

In other words two genes are linked in the resulting graph if
they have the possibility to catalyze two successive reactions, the main
product of the first one being the main substrate of the second
one. Although it is far from being certain that all the genes candidates
to catalyze a given reaction (because they are assigned an EC number
supposed to represent a family of potential enzymes catalyzing the
reaction) actually catalyze it in the cell, these data nevertheless
provide a global picture of the possible relationships between genes in
terms of catalyzing properties. In particular a path in this graph
corresponds to a possible series of reactions catalyzed by the
successive genes met along the path.

The resulting graph involves 774 genes of {\em S. Cerevisiae}, linked with
16,650 edges.

\subsection{Microarray data}
Publicly available microarray expression data were collected from the
Stanford Microarray Database \cite{sher01}. The data include yeast
response to various experimental conditions, including metabolic shift from
fermentation to respiration  \cite{deri97}, alpha-factor block release, cdc15
block release, elutriation time course, cyclin over-expression \cite{spel98}, sporulation \cite{chu98}, adaptive evolution
\cite{fere99}, stress response \cite{gasc00}, manipulation in
phosphate level \cite{ogaw00}, cell cycle \cite{zhu00}, growth
conditions of excess copper or copper deficiency \cite{gros00}, DNA
damage response \cite{gasc01}, 
and  transfer from a fermentable to a nonfermentable carbon source
\cite{kuhn01}.

Combining these data results in 330
data points available for 6075 genes, i.e., almost all known or
predicted genes of {\em S. cerevisiae}.
Each data point produced by a DNA microarray hybridation experiment
represents the ratio of expression levels of a particular gene under two
experimental conditions. Following \cite{eise98,brow00} we don't work directly
with this ratio but rather with its normalized logarithm defined as:
$$
\forall (x,i) \in \XX \times \{1,\ldots,330\}, \quad e(x)_i = \frac{\log
E_{x,i} / R_{x,i}}{\sqrt{\sum_{j=1}^{330} \log^2
E_{x,i} / R_{x,i}}},
$$
where $E_{x,i}$ is the expression level of gene $x$ in experiment $i$
and $R_i$ is the expression level in the corresponding reference state.
Missing values were estimated
with the software KNNimpute \cite{troy01}.

\subsection{Functional classes}
The January 10, 2002, version of the functional classification catalogue of
the Comprehensive Yeast 
Genome Database (CYGD) \cite{mewe02} is a comprehensive classification
of 3936 yeast genes into 259 functional classes organized in a hierarchy.
The classes vary in size between 1 and 2258 genes (for the class 
``subcellular localization''), and not all of them are supposed to be
correlated with gene expression \cite{brow00}. Only classes with at
least 20 genes (after removing the genes present in the gene graph, see
next Section) are considered as benchmark datasets for function
prediction algorithm in the sequel, which amounts to 115 categories.

\subsection{Gene function prediction}

Following the general approach presented in Section \ref{sec:extpro} the
gene prediction experiment involves two steps:
\begin{itemize}
\item The 669 genes in the gene graph derived from the pathway database with
known expression profiles are used to perform the feature extraction
process by solving (\ref{eqn:eigpro}). 
\item The resulting linear features are extracted from the expression
      profiles of the disjoint  set of 2688 genes which
are in the CYGD functional catalogue but not in the pathway
      database. Systematic evaluation of the performance of support
      vector machines to predict each CYGD class either from the
      expression profiles themselves \cite{brow00} or from the features
      extracted is then performed on this set of genes using 3-fold
      cross-validation averaged over 10 iterations.
\end{itemize}

Support vector machine (SVM) \cite{vapn98,cris00,scho02} is a class
of machine learning algorithms for supervised classification which has
been shown to perform better that other machine learning techniques,
including  Fisher's linear discriminant, Parzen windows and decision
trees on the problem of gene functional classification from expression
profiles \cite{brow00}. We therefore use SVM as a state-of-the-art
learning algorithm to assess the gain resulting from replacing the
original expression profiles by vectors of features.

Experiments were carried out with SVM Light \cite{joac99}, a public and
free implementation of SVMs. 
To ensure a comparison as fair as possible between different data
representations, all vectors were scaled to unit length before being sent
to the SVM, and all SVM used
a radial basis kernel with unit width, i.e., $k(x,y) =
\exp(-||x-y||^2)$. The trade-off parameter between training error and
margin was set to its default value (namely 1 in the case where all
vectors have unit length), and the cost factor by which training errors
on positive examples outweigh errors on negative examples was set equal
to the ratio of the number of positive examples and the number of
negative examples in the training set.

We compared the performance of SVM working directly on the expression
profiles. as in \cite{brow00}, with SVM working on the vectors of
features extracted by the procedure described in this paper, for various
choices of regularization parameters $\delta$, width of the diffusion
kernel $\tau$ and numbers of features selected.

For each experiment the performance is measured by the ROC index,
defined as
the area under the ROC curve, i.e., the plot of true positives versus
false positives, and
normalized to 100 for a perfect classifier. The ROC curve itself is
obtained by varying a threshold and classify genes by comparing the
score output by the SVM with this threshold. A random classifier has an average
ROC index of 50.

\subsection{Setting the parameters}
Our feature extraction process contains two free parameters, namely the
width $\tau$ of the diffusion kernel and the regularization parameter
$\delta$. Intuitively, the larger $\tau$ and $\delta$, the smoother and
more relevant the features extracted, at the expense of a decrease
between their correlations. As pointed out in \cite{bach01} the
parameter $\delta$ is expected to decrease linearly with $n$, and a
reasonable value is $\delta = 0.001$ for $n$ of the order of 1000. An
initial value of $\tau=1$ was chosen.

We varied independently $\delta$ and $\tau$ in order to check their
influence. For a fixed $\delta = 0.001$ we tested the performance of
SVM based on the features extracted with the parameter $\tau \in \{0.5 ,
1 , 2 , 5\}$, where all 330 features are used. Table \ref{tab:tau}
shows the ROC index averaged over all 115 classes with more than 20
genes for each of the four SVM,
as well as the percentage of classes best predicted by each method.
The best performance is reached for $\tau = 1$, with an important
deterioration when $\tau$ increases to $5$. A larger $\tau$ means by
(\ref{eqn:normdiff}) that rugged features are more strongly penalized, so
larger $\tau$ tend to generate smoother features. The deterioration when
$\tau$ increases shows the importance of not excessively penalizing
ruggedness.
\begin{table}
\center
\caption{Performance comparison for various $\tau$}\label{tab:tau}
\begin{tabular}{cccc} \hline
$\delta$ & $\tau$ & Average ROC & Percentage of classes best predicted \\ \hline
0.001 & 0.5 & 61.4 & 37 \\
0.001 & 1 & 61.4 & 35 \\
0.001 & 2 & 60.0 & 20 \\
0.001 & 5 & 55.2 & 8 \\
\hline
\end{tabular}
\end{table}

We also checked the influence of the regularization parameter $\delta$,
which controls the trade-off between correlation on the one hand,
smoothness and relevance on the other hand. Table \ref{tab:delta}
compares the performances of SVM based on the features extracted with
the parameters $\tau = 1$ and $\delta \in \{0.0005 , 0.001 , 0.002 , 0.005\}$.
This shows a small (in terms of ROC index increase) but consistent (in
terms of number of classes best predicted) increase in performance when
$\delta$ increases 
from $0.0005$ to $0.005$. This illustrates the importance of
regularization, and therefore the improvement gained by imposing some
smoothness and relevance constraints to the features.
\begin{table}
\center
\caption{Performance comparison for various $\delta$}\label{tab:delta}
\begin{tabular}{cccc} \hline
$\delta$ & $\tau$ & Average ROC & Percentage of classes best predicted \\ \hline
0.0005 & 1 & 61.4 & 17 \\
0.001 & 1 & 61.4 & 18 \\
0.002 & 1 & 61.4 & 25 \\
0.005 & 1 & 61.6 & 39 \\
\hline
\end{tabular}
\end{table}

\subsection{Number of features}
\begin{figure}\label{fig:nbfeat}
\centerline{\includegraphics[width=28pc]{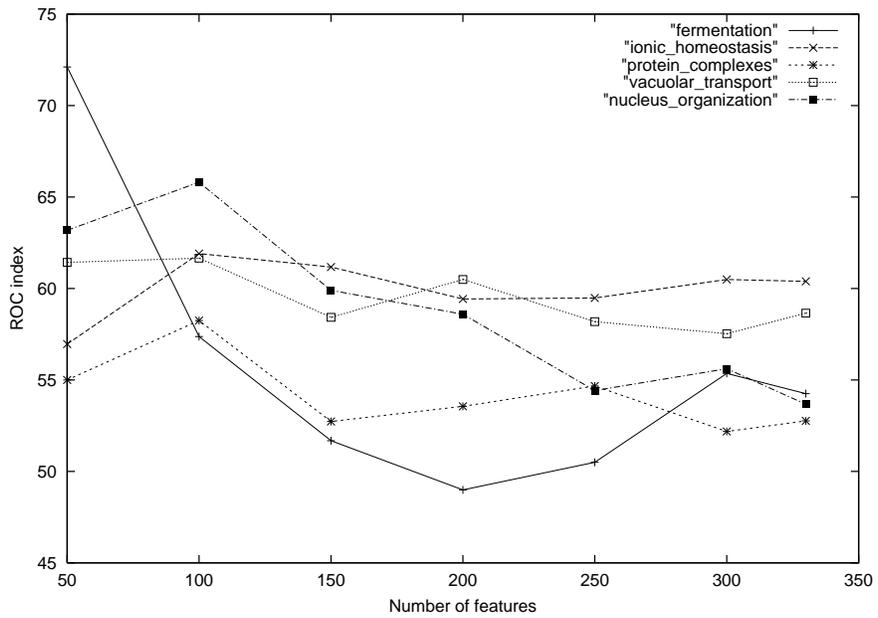}}
\caption{Classification performance for various classes}
\end{figure}

From now on we fix the parameters to $\tau = 1$ and $\delta = 0.001$. As
the feature extraction process is supposed to extract up to $p=330$
features by decreasing biological relevance, one might ask if
classification performance could increase by only keeping the most
relevant features, and hopefully removing noise by discarding the
remaining ones. To check this we measured the performance of SVM using
an increasing number of features. Results are shown on Table
\ref{tab:nb}, and show that it is on average more interesting to use all
features as the performance increases with the number of features
used. Exceptions to this average principle include classes such as
fermentation, ionic homeostasis, assembly of protein complexes, vacuolar
transport, phosphate metabolism or nucleus organization, which are
better predicted with less than 100 features as shown on Figure \ref{fig:nbfeat}
\begin{table}
\caption{Performance comparison for various numbers of features, with
 $\delta=0.001$ and $\tau=1$}\label{tab:nb}
\center
\begin{tabular}{ccc} \hline
Number of features & Average ROC & Percentage of classes best predicted \\ \hline
50 & 55.3 & 3 \\
100 & 57.9 & 10 \\
150 & 58.9 & 9 \\
200 & 59.9 & 7 \\
250 & 60.6 & 17 \\
300 & 61.2 & 17 \\
330 & 61.4 & 37 \\
\hline
\end{tabular}
\end{table}

\subsection{Functional classification performance}
\begin{figure}\label{fig:perf}
\centerline{\includegraphics[width=28pc]{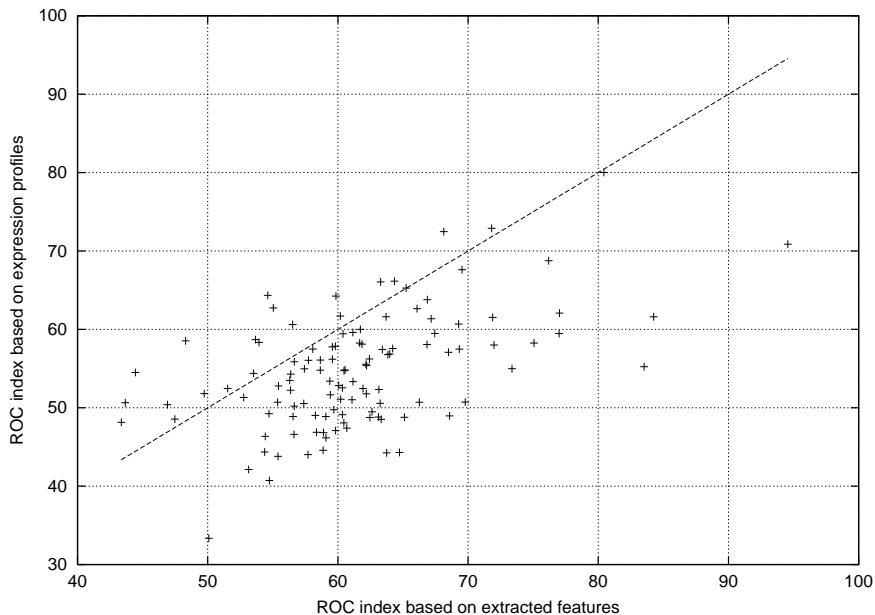}}
\caption{Comparison of the classification performance of SVM based on
 expression profiles (y axis) or extracted features (x axis). Each point
 represents one functional class.}
\end{figure}

In order to check whether the features extraction provides any advantage
over the direct use of expression profiles for gene function prediction
we finally compared the performance of a SVM using all
features extracted with the parameters $\delta=0.001$ and $\tau=1$, with
the performance of a SVM using directly the gene expression
profiles. Figure \ref{fig:perf} shows the ROC index obtained by each of
the two methods for all 115 functional classes. Except for a few
classes, there is a clear improvement in classification performance when
the genes are represented as vectors of features, and not directly as
expression profiles. 

Table \ref{tab:perf} shows that the ROC index averaged over all
 classes increases significantly between the two representations (from
 54.9 to 61.2). Moreover Figure \ref{fig:perf} shows that most of
 the classes seem almost impossible to learn from their expression
 profiles only (when the ROC index is around 45 - 55, i.e. not better than
 a random classifier), but can somehow be learned by their vectors of
 features, as the ROC index jumps in the range 55-65 for many of those
 classes. Some classes exhibit a dramatic increase in ROC index, as shown
 in Table \ref{tab:perf2} which lists the classes largest absolute
 increase in ROC index between the two experiments.
\begin{table}
\caption{ROC index averaged over 115 functional classes by SVM using
 different representations of the data}\label{tab:perf}
\center
\begin{tabular}{cc} \hline
Data representation & Average ROC \\ \hline
Expression profiles & 54.6 \\
Vector of features & 61.4 \\
\hline
\end{tabular}
\end{table}

\begin{table}
\caption{ROC index for the prediction of categories based on expression
 profiles or features vectors. The categories listed are the one which
 exhibit the largest increase in ROC index between these two
 representations.}\label{tab:perf2}
\center
\begin{tabular}{cccc} \hline
Class & Expression & Features & Increase\\ \hline
Heavy metal ion transporters (Cu, Fe, etc.) & 55.2 & 83.5 & +28.3\\
Ribosome biogenesis & 70.9 & 94.6 & +23.7\\
Protein synthesis & 61.6 & 84.3 & +22.7 \\
Directional cell growth (morphogenesis) & 44.3 & 64.7 & +20.4 \\
Regulation of nitrogen and sulphur utilization & 49.0 & 68.6 & +19.6\\
Nitrogen and sulfur metabolism & 44.3 & 63.8 & +19.5 \\
Translation & 50.7 & 69.8 & +19.1\\
Cytoplasm & 55.0 & 73.4 & +18.4 \\
Endoplasmic reticulum & 59.5 & 77.0 & +17.5\\
Amino acid transport & 75.1 & 58.3 & +16.8 \\
\hline
\end{tabular}
\end{table}

\section{Discussion and conclusion}\label{sec:discussion}

This paper proposes an algorithm to extract features from gene expression
profiles based on the knowledge of a biochemical network linking a
subset of
genes. Based on the simple idea that relevant features are likely to
exhibit correlation with respect to the topology of the network, we
end up with a formulation which involves encoding the network and the
set of expression profiles into to kernel functions, and performing a
regularized canonical correlation analysis in the corresponding
reproducible kernel Hilbert spaces.

Results presented in Section \ref{sec:experiment} are encouraging and
confirm the intuition that incorporating valuable information, such as
the knowledge of the precise position of many genes in a biochemical
network, helps extracting relevant informations from expression
profiles. While this problem has still attracted relatively few attention
because the number of expression data has always been small compared to
the number of genes until recently, it is expected to be more and more important as the
production of expression data becomes cheaper and the underlying
technology more widespread.

A detailed analysis of the experimental results reveals that
functional categories related to metabolism, protein synthesis and subcellular
localization benefit the most from the representation
of genes as vectors of features. In the case of metabolism and protein synthesis
related categories, this can be explained by the fact that many pathways
related to this process are present in the pathway database, so relevant
features have probably been extracted. The case of subcellular
localization proteins is more surprising, as they seem to be more related to
structural properties than functional properties of the genes, but
certainly reflects the functional role of the organelles themselves. As
an example a sudden need of energy might promote the activity in
mitochondria and require the synthesis of proteins to be directed to
this location, even though they might not be directly involved as
enzymes.

On the technical point of view the approach developed in this paper can
be seen as an attempt to encode various types of information about genes
into kernels. The diffusion kernel $K_1$ encodes the
gene network, and the linear kernel $K_2$ summarizes the expression
profiles. Recent research shows that this approach can in fact be
generalized to many other sources of information about genes, as many
kernels have been engineered and continue to be developed for
particular types of data. Apart from classical kernels for
finite-dimensional real-valued vectors
\cite{vapn98} which can be used to encode any vectorial gene
representation, e.g. expression profiles, and from diffusion kernels
which can encode any gene network, e.g. network derived from biochemical
pathway or protein interaction networks, relevant examples of recently
developed kernels include the Fisher kernel to encode how the
amino-acid sequence of a protein is related to a given hidden Markov
model \cite{jaak00} or to encode the arrangement of transcription factor
binding site motifs 
in its promoter region \cite{pavl01}, several string kernels to
encode the information present in the amino-acid sequence itself
\cite{haus99,watk00,lesl02,vert02,lodh02}, or a tree kernel to encode
the phylogenetic profile of a protein \cite{vert02b}. This increasing
list suggests a unified framework to represent various types of
informations, which is obtained by ``kernelizing the
proteome'', i.e., tranforming any type of information into an adequate kernel.

Parallel to the apparition of new kernels recent years have witnessed
the development of new methods, globally referred to as {\em kernel
methods}, to perform various data mining algorithm from the
knowledge of the kernel matrix only. Apart from the most famous
support vector machine algorithm for classification and regression
\cite{bose92,vapn98}, other kernel methods include principal component
analysis \cite{scho99}, clustering \cite{benh01}, Fisher discriminants
\cite{mika99} or independent component analysis \cite{bach01}. 

These recent developments
open the door to new analysis opportunities which we believe can be
particularly suited to the new discipline of proteomics whose central
concepts, genes or proteins, are defined through a variety of different
points of view (as sequences, structures, expression patterns, position
in networks, ...), the integration of which promises to unravel some of
the secrets of life.

\section{Acknowledgements}
We would like to thank Yasushi Okuno for help and advices with the
 pathway data, and Olivier Bousquet for simulating discussions. This work was
 supported by the Research for the Future Program of the Ministry of
 Education, Culture, Sports, Science and Technology, Japan.

\bibliography{/mnt/export/home/vert/html/bibli2/bibli.bib}
\end{document}